\documentclass[aps,pra,showpacs,twocolumn,superscriptaddress]{revtex4}
\usepackage{psgo,array}
\setgounit{0.4cm}
\usepackage{graphicx}

\begin{document} 
\title{Move ordering and communities in complex networks describing the game of go}

\author{V. Kandiah}
\affiliation{%
Universit\'e de Toulouse; UPS; Laboratoire de
 Physique Th\'eorique (IRSAMC); F-31062 Toulouse, France
}
\affiliation{CNRS; LPT (IRSAMC); F-31062 Toulouse, France}
\author{B. Georgeot}
\affiliation{%
Universit\'e de Toulouse; UPS; Laboratoire de
 Physique Th\'eorique (IRSAMC); F-31062 Toulouse, France
}
\affiliation{CNRS; LPT (IRSAMC); F-31062 Toulouse, France}
\author{O. Giraud}
\affiliation{LPTMS, CNRS and Universit\'e Paris-Sud, UMR 8626, B\^at. 100,
91405 Orsay, France}

\begin{abstract}
We analyze the game of go from the point of view of complex networks. We construct three different directed networks of increasing complexity, defining nodes as local patterns on plaquettes of increasing sizes, and links as actual successions of these patterns in databases of real games.
We discuss the peculiarities of these networks compared to other types of networks.  We explore the ranking vectors and community structure of the networks and show that this approach enables to extract groups of moves with common strategic properties. 
We also investigate different networks built from games with players of different levels or from different phases of the game. We discuss how the study of the community structure of these networks may help to improve the computer simulations of the game. More generally, we believe such studies may help to improve the understanding of human decision process.  
\end{abstract}

%01.80.+b       Physics of games and sports
%89.75.Da       Systems obeying scaling laws
%89.20.-a       Interdisciplinary applications of physics
%89.75.Hc       Networks and genealogical trees 
%\date{\today}
\date{May 23, 2014}
\maketitle

\section{Introduction}
The study of complex networks has become more and more important in the recent past. In particular, communication and information networks have become ubiquitous in everyday life. New tools have been created to understand the mechanisms of growth of such networks and their generic properties. On the other hand, it has been realized that other phenomena can also be modelized by such tools, e.g. in social sciences, linguistics, and so on \cite{barabasi}.

However, the tools of complex networks were never applied to the study of human games. Nevertheless, games represent one of the oldest human activities, and may give insight into the  human decision-making processes. In \cite{gonet}, a network was built that describes the game of go, one of the oldest and most famous board games. The complexity of the game is such that no computer program has been able to beat a good player, in contrast with chess where world champions have been bested by game simulators.  It is partly due to the fact that  
the total number of possible allowed positions in go is about $10^{171}$, compared to e.g. only $10^{50}$  for chess \cite{TroFar07}. In fact, among traditional board games it has by far the largest state space complexity \cite{Herik02}. Part of the complexity of the game of go comes from this large number of different board states, due to the fact that it is played on a board (the goban) composed of 19 vertical lines and 19 horizontal lines, implying 361 possible positions, against 64 in chess. Also, it is very hard for a computer to evaluate the positional advantages in the course of the game, while in chess the capture of different pieces can be easily compared. 

Due to that, the study of computer go has become an important subfield of computer science. Its main challenge is to estimate a value function of moves, that is, a function which assigns a value to each move, given a certain state of the goban. Traditional approaches evaluate the value function by using huge databases of patterns, from initial patterns to life-and-death situations, and can learn to predict the value of moves by reinforcement learning (see e.g.~\cite{Schrau}). By contrast, the recently introduced Monte-Carlo go does not rest primarily on expert knowledge. Its basic principle is to evaluate the value of a move by playing at random, according to the rules of go, from a given state, until the end, so that a value can be assigned to the move. Playing thousands of games allows to estimate the value function for each move. This approach has proved way more efficient than the classical approaches \cite{MonteCarloGo,Brown12}.

Many improvements have since then been included in Monte-Carlo go. In particular, Monte-Carlo tree search, implemented in computer programs such as Crazy Stone \cite{Cou07} or MoGo \cite{Mogo}, is based on the construction of a tree of goban states, where new states are added iteratively as they are met in a simulation. The value function is updated depending on the outcome of each randomly played game. Random moves are chosen according to some playing policy which can itself be biased towards certain moves (for instance, capture whenever possible), and in such a way that most promising moves are more carefully explored, but with an incentive to visit moves with a large uncertainty on their actual value. Recent improvements allow to improve the exploration of the tree \cite{GelKoc12}. To get faster estimates of the value of a move the RAVE (Rapid Action Value Estimation) algorithm, or its Monte-Carlo version, attributes to a move in a given state $s$ the average outcome of all games where that move is played {\it after} state $s$ has been encountered \cite{GelSil11}. In order to account for rarely visited states, a heuristic prior knowledge can be fed into the algorithm to attribute an a priori value to a move, such as e.g. the value of its grand-father, or a value depending on local patterns \cite{GelSil07}.

Although global features, such as chain connections, or the influence of stones over domains of the goban, are crucial in the game of go, local features can be used at many places in the algorithms of computer go, for instance to improve the heuristic value function which initializes the value of each move, or to get a faster estimate of the exact value \cite{BouzyChaslot,Elo}.

There is therefore a clear interest in having a better understanding of local features in the game of go.  
In \cite{gonet}, two of us introduced a small network based on local positional patterns and showed that it can be used to extract information on the tactical sequences used in real games. However, the small size of the plaquettes made it difficult to disambiguate many strategically different moves. In the present paper, we construct three networks based on positional patterns of different sizes, and study their properties. The network size varies by a factor one hundred, and the largest one enables to specify more precise features that were difficult to disambiguate in \cite{gonet}. In particular, the community structure is much easier to characterize and discuss.  After presenting the details of the construction of the networks (Section II) we study their global properties such as ranking vectors and spectra of the Google matrix, contrast them to other types of networks,  and relate them to specific features of the game (Section III). In Section IV, we study in detail the characterization of communities of nodes in the networks, a well-known subject in network theory, which in our case enables to regroup tactical moves with common features. In Section V we propose the construction of different networks corresponding to specific phases of the game or to different levels of players.

\section{The go networks}

The game of go is played on a board (goban) of $ 19 \times 19$ intersections of vertical and horizontal lines.  Each player  alternately places a stone of his/her color (black or white) at an empty intersection. Empty intersections next to a group of connected stones of the same color are called ''liberties".  If only one liberty remains, the group of stones is said to be in atari. When the last liberty is occupied and the group is entirely surrounded by the opponent, its stones must be removed. The aim of the game is to surround large territories and to secure their possession.  Good players follow general strategies through a series of local tactical fights.  We construct the networks representing the game by connecting local moves played in the same neighbourhood (note the similarity with some language networks \cite{language} which are also based on local features). We describe a move by identifying the empty intersection $(h,v)$ (with $1\leq h,v\leq 19$) where the new stone is placed.

The vertices of our networks are based on what we  call ''plaquettes'', i. e. a part of the goban with a given shape and size which depends on the network.   Each plaquette corresponds to a certain pattern of white and black stones with an empty intersection at its center, on which black will put a stone. We identify plaquettes which are related by translation on the goban or by a symmetry of the square, and additionally those with colors swapped.  

The first network we consider (Network I) is made as in \cite{gonet} by taking as plaquettes 
squares of 3 $\times$ 3 intersections, which are subparts of the goban of the form  $\{(h+r,v+s),-1\leq r,s\leq 1\}$ (edges and corners of the board can be accounted for by imagining additional dummy lines outside the board).  Once borders and symmetries are taken into account, we obtain as vertices of network I a total of 1107 nonequivalent plaquettes (with empty centers).

Network II is made by also taking squares of 3 $\times$ 3 intersections and identifying plaquettes related by symmetry, but we also include the atari status of the four nearest-neighbour points from
the center. Atari status assesses if the chain of stones to which a given stone belongs has only one liberty (one empty intersection connected to it). Removing the last liberty of a chain in atari entails the capture of the whole group. In this case, many seemingly possible configurations are not legal 
since they would contradict the atari status. This leaves 2051
legal nonequivalent plaquettes with empty centers (the same figure was found in \cite{Coulom2}).

Network III is based on diamond-shape plaquettes: the 3 $\times$ 3 plaquettes discussed above plus
the four at distance two from the center in the four directions left, right, top, down.
We still identify plaquettes related by symmetry, but do not take into account the atari status.
This gives us 193995 nonequivalent plaquettes with empty centers, which are the vertices of network III (96771 are so rare that they are actually never used in our database of games).

\begin{figure}[!h]
\begin{center}
\includegraphics[trim=0.00 0.00 0.00 0.00,clip, width=\columnwidth]{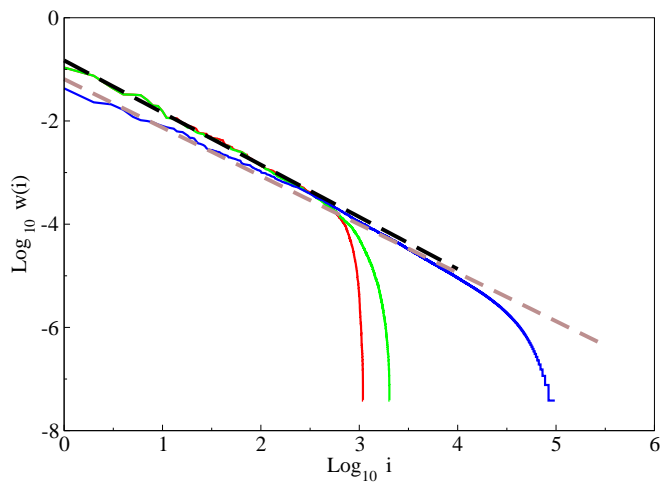}
\end{center}
\caption{(Color online) Distribution of frequency of occurrences $w(i)$ of different plaquettes for the three different networks (full lines), from left to right at the bottom: red: square plaquettes (network I), green: square plaquettes with atari status (network II), blue: diamond plaquettes (network III)(see text)(data from networks I and II are indistinguishable over parts of the curves). The dashed straight lines are power law fits with slopes $-1.02$ (black upper line, fit of network II) and $-0.94$ (brown lower line, fit of network III).  \label{Zipf}}
\end{figure}

We have identified the occurrence of these different plaquettes in games from
a database available at \cite{database}. This database contains the sequence of moves of 135663 different games corresponding to players of diverse levels (the level 
of the players is marked by a number of dans, from 1 to 9). The games recorded have been played online, and the dans have been mutually assessed according to the results of these plays.
The frequency of the different plaquettes is shown in Fig.~\ref{Zipf}. It can be compared to Zipf's law,  an empirical law seen in many natural distributions (word frequency, city sizes, chess openings...) \cite{Zipf1,Zipf2,Zipf3,Zipf4}. For items ranked according to their frequency, it corresponds to a power-law decay of the frequency versus  the rank. The data presented in Fig.~\ref{Zipf} show that the three different network choices all give rise to a distribution following Zipf's law, although the slope varies from $\approx -1$ (networks I and II) to a slightly slower decay for the largest network (network III).

\begin{figure}[!h]
\begin{center}
\includegraphics[trim=0cm 0cm 0cm 0.0cm,clip,width=0.9\columnwidth]{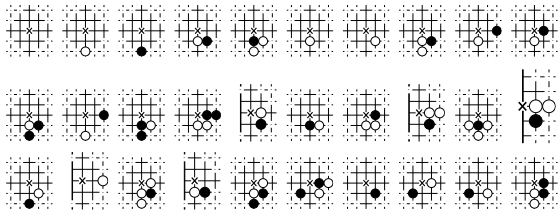}
\end{center}
\caption{Top 30 plaquettes in frequency of occurrences for the network III (diamond plaquettes). Black plays at the black cross. Dotted intersections are outside the diamond plaquette and their status is unknown. \label{plaqueZ}}
\end{figure}

We display in Fig.~\ref{plaqueZ} the top 30 moves in order of decreasing frequency of occurrences for network III. The most common correspond to few stones on the plaquettes, which is natural since these ones are present at the beginning of almost all local fights, while the subsequent moves differ from games to games.

To define links of our three networks, we connect vertices corresponding to moves $a$ and $b$ played at $(h_a,v_a)$ and $(h_b,v_b)$ on the board if $b$ follows $a$ in a game of the database and $\max\{|h_b-h_a|,|v_b-v_a|\}\leq d$ where $d$ is some distance. Here contrary to \cite{gonet} we put a link only between $a$ an the first move following $a$ in the specified zone. Each integer $d$ corresponds to a different network. It specifies the distance beyond which two moves are considered unrelated.  In \cite{gonet}, different values of $d$ were considered and it was shown that the value $d=4$ was the most relevant, allowing a correct hierarchization of moves: related local fights are kept while far away tactical moves are not taken into account.  In the following we will thus retain this value $d=4$. Two vertices are thus connected by a number of directed links given by  the number of times the two corresponding moves follow each other in the same neighbourhood of the goban in the games of the database.

\begin{figure}[!h]
\begin{center}
\includegraphics[trim=0.00 0.00 0.00 0.00,clip,width=\columnwidth]{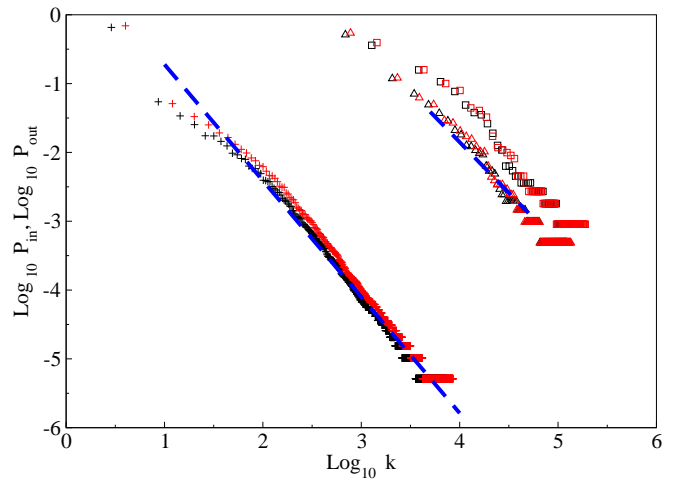}
\end{center}
\caption{ (Color online) Distribution of incoming links $P_{\textrm{in}}$ (black) and outgoing links $P_{\textrm{out}}$ (red/grey) for the three different networks; square plaquettes (network I) (squares), square plaquettes with atari (network II) (triangles), diamond plaquettes (network III) (crosses). The dashed lines are power law fits with slopes $-1.47$ (right) and $-1.69$ (left). \label{links}}
\end{figure}

With this definition, the three networks are now defined, with vertices connected by directed links. The total number of links including degeneracies is 26116006 links. The numbers without degeneracies are respectively 558190 (network I), 852578 (network II) and 7405395 (network III). The link distributions are shown in Fig.~\ref{links}; it is close to a power-law. This implies that the networks present the
scale-free property \cite{barabasi}. One can notice a symmetry between ingoing and outgoing
links, which  is a peculiarity of this problem, and is not seen in e.g. the World Wide web, where the exponent for $P_{\textrm{out}}$ ($\approx -2.7$) is different from the one for $P_{\textrm{in}}$ ($\approx -2.1$) \cite{donato}.  Here exponents are similar and close to $1.5$, intermediate between these two values. Our results indicate the presence of a symmetry (at least at a statistical level) between moves that follow many different others and moves which
have many possible followers.  This symmetry is natural, since in many cases (i.e. in the course of a local fight) the occurrence of a plaquette in the database implies the presence of both an ingoing and an outgoing link.

\section{Ranking vectors and spectra of Google matrices}

We have presented up to now the construction  of our networks for the game of go, and their global statistical properties. To get more insight into the organization of the game, we use tools developed in the framework of network theory, in order to hierarchize vertices of a network. Such tools are routinely used by search engines to decide in which order answers to queries are presented.  The general strategy is to build a ranking vector, whose value on each vertex will measure its importance. A  famous vector of this type is the PageRank
\cite{brin,googlebook}, which has been at the basis of the Google search engine. It can be obtained
from the Google matrix $G$, defined as $G_{ij}=\alpha S_{ij}+(1-\alpha)\, ^{t}ee/N$, where $e=(1,...,1)$, $N$ is the size of the network, $\alpha$ is a parameter such that $0<\alpha\leq 1$ (we chose $\alpha=1$ in the computations in this paper), and
$S$ is the weighted adjacency matrix.  The latter starts from the adjacency matrix where the value of the entry $(i,j)$ corresponds to the number of links from vertex $j$ to vertex $i$; then one replaces any column of 0 by a column of $1$, and one normalizes the sum of each column to 1. This ensures that the matrix $G$ has the mathematical property of stochasticity. The PageRank vector is defined as the right eigenvector of the matrix $G$ associated with the largest eigenvalue $\lambda=1$. It singles out as important vertices the ones with many incoming links from other important nodes. Equivalently, it can be seen as giving the average time a random surfer on the network will spend on each vertex. Indeed, the process of iterating $G$ can be seen as the action of a random surfer choosing randomly at each node to follow a link to another node. The largest eigenvalue corresponds to the equilibrium distribution of the surfer, and gives the average time spent on each node.  Other ranking vectors which can be built from the matrix $G$ include
the CheiRank vector \cite{dima}, and the Hubs and Authorities of the HITS algorithm \cite{hits}. While PageRanks and Hubs attribute importance to vertices depending on their incoming links,
CheiRanks and Authorities stem from outgoing links.  In particular, CheiRank can be defined as the PageRank of the ``dual'' network where all links are inverted. We denote the Google matrix of this dual network by $G^*$. 

\begin{figure}[!h]
\begin{center}
\includegraphics[trim=0.00 0.00 0.00 0.00,clip,width=\columnwidth]{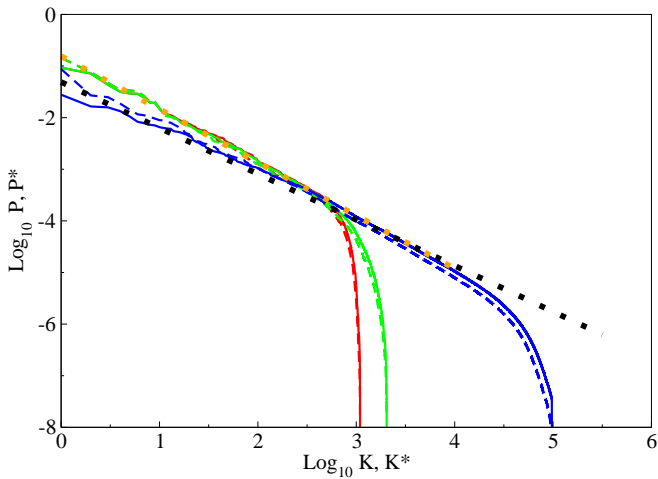}
\end{center}
\caption{(Color online) Distribution of ranking vectors (normalized by $ \sum_K P(K)=\sum_{K^*}P^*(K^*)=1$) for the three different networks: PageRank $P(K)$ (solid lines) and CheiRank $P^*(K^*)$ (dashed lines), same color code for the networks as in Fig.~\ref{Zipf} (data from networks I and II are indistinguishable over parts of the curves).  The dotted lines are power law fits with slopes $-1.03$ (orange upper line, fit of network II) and $-0.89$ (black lower line, fit of network III). \label{PR} }
\end{figure}

In Fig.~\ref{PR} the distributions of PageRank and CheiRank are shown for the three networks , showing that 
ranking vectors follow an algebraic law, with a slightly different exponent for the largest network.
Similarly as for the link distribution, one sees a symmetry between distributions of ranking vectors based on ingoing links and outgoing links, again an original feature which can be related to the statistical symmetry between ingoing and outgoing links. 

\begin{figure}[!h]
\begin{center}
\includegraphics[trim=0cm 0cm 0cm 0.0cm,clip,width=0.9\columnwidth]{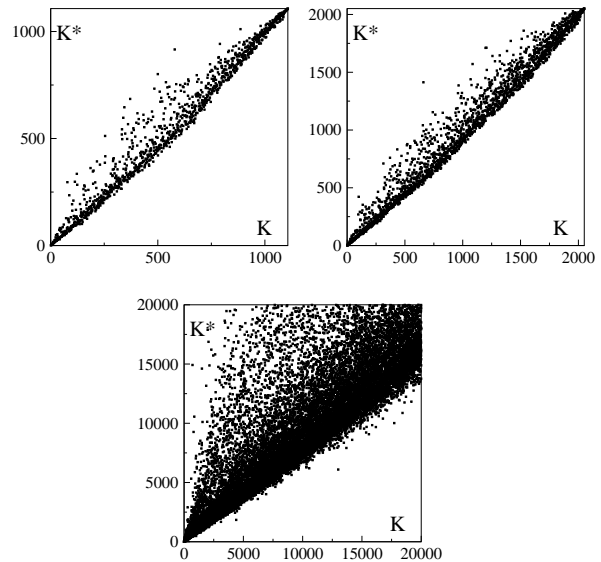}
\end{center}
\caption{PageRank-CheiRank correlation plot of the three different networks : square plaquettes (network I)(top left), square plaquettes with atari status (network II)(top right) and diamond plaquettes (network III)(bottom). PageRank $K$ is given in $x$-axis and CheiRank $K^*$ in $y$-axis, the plot of network III is a zoom on the top $20000$ moves in both $K$ and $K^*$. \label{corr}}
\end{figure}

In order to check to what extent this symmetry affects the ranking vectors, we plot in  Fig.~\ref{corr} the CheiRank $K^*$ as a function of the PageRank $K$. It indeed shows that the two quantities are not independent, and strong correlations
between PageRank and CheiRank do exist. This symmetry is not visible in general for other networks (see e.~g.~\cite{trade} where similar plots are shown in the context of world trade, displaying much less correlation).  Nevertheless, the symmetry is clearly not exact, especially for the largest network (a perfect correlation will produce points only on the diagonal); the plots are not even symmetric with respect to the diagonal.  Thus PageRank and CheiRank produce genuinely different information on the network.

\begin{figure}[!h]
\begin{center}
\includegraphics[trim=0cm 0cm 0cm 0.0cm,clip,width=0.9\columnwidth]{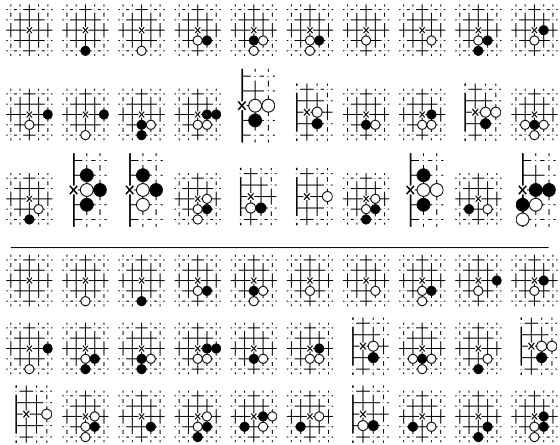}
\end{center}
\caption{Top 30 plaquettes for first eigenvector of $G$ (PageRank)(top) and $G^*$ (CheiRank)(bottom) of the network III. \label{plaque}}
\end{figure}

Fig.~\ref{plaque} shows the first 30 plaquettes in decreasing importance in the PageRank and CheiRank vectors. The correlation between the two sequences is clearly visible, although it is again not perfect. We note that these sequences are also very similar to the one obtained by just counting the move frequency (as in Zipf's law): most frequent moves tend to dominate the ranking vectors. 

\begin{figure}[!h]
\begin{center}
\includegraphics[trim=0cm 0cm 0cm 0.0cm,clip,width=0.9\columnwidth]{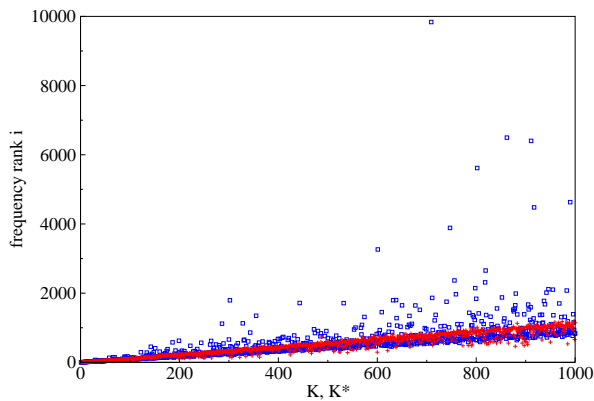}
\end{center}
\caption{Correlation plot of PageRank-CheiRank vs frequency of moves for network III (diamond plaquettes) (only first 1000 moves in $K$ are shown); blue squares: PageRank $K$, red crosses: CheiRank $K^*$.  \label{corrZ}}
\end{figure}

However, as Fig.~\ref{corrZ} shows, the correlation between ranking vectors and frequency ordering is far from perfect, especially for the PageRank, which can be extremely different from the rank obtained by frequency. This shows that the ranking vectors present an information obtained from the network construction, which differs from the mere frequency count of moves in the database.  Indeed, as explained above the frequency count is related to the link distribution due to the construction process of the network.  It is known in general that the PageRank has some relation with the distribution of ingoing links, but with the significant difference that it highlights nodes whose ingoing links come from (recursively defined) other important nodes.  This was the basis of the fortune of Google and in our case means that highlighted moves correspond to plaquettes with ingoing links coming from other important plaquettes. Thus the PageRank underlines moves to which converge many well-trodden paths of history in the different games of the database. The CheiRank does the same in the reverse direction, highlighting moves which open many such paths.

\begin{figure}[!h]
\begin{center}
\includegraphics[trim=0cm 0cm 0cm 0.0cm,clip,width=0.9\columnwidth]{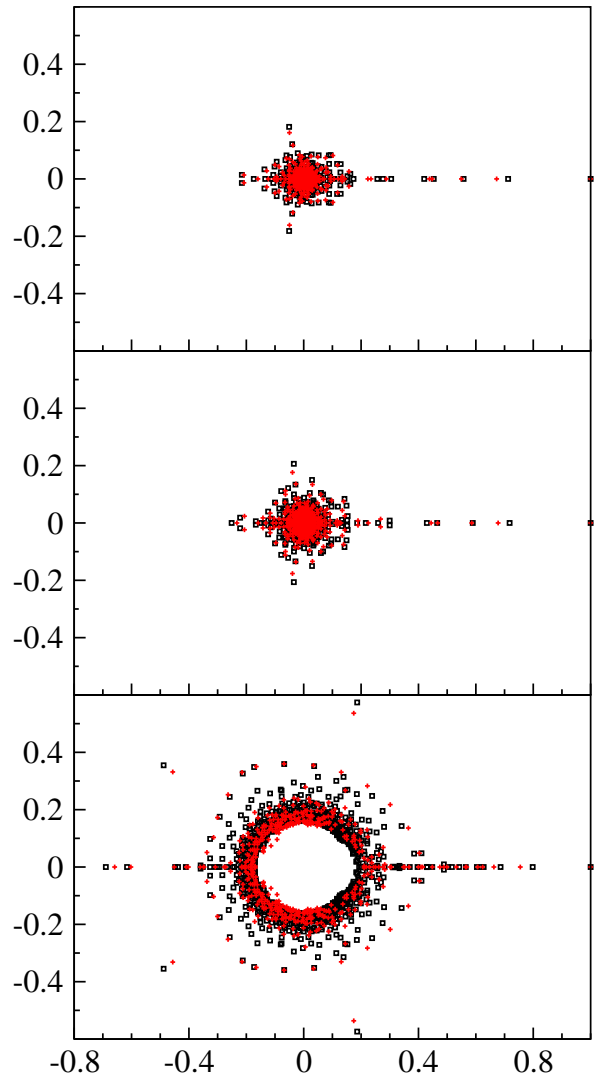}
\end{center}
\caption{(Color online) Spectrum in the complex plane of $G$ (black squares) and $G^*$ (red/grey crosses) for the three different networks : I (top), II (middle) and III (bottom). \label{spectrum}}
\end{figure}

The ranking vectors discussed above are just one eigenvector of the matrices associated with a given network. However, other eigenvalues and their associated eigenvectors also contain information about the network. We have computed the spectrum of the Google matrix for the three networks; they are shown in Fig.~\ref{spectrum}.  For square plaquettes (network I) and square plaquettes plus atari status (network II) all eigenvalues are computed. In the case of the largest network, standard diagonalization techniques could not be used and therefore we used an Arnoldi-type algorithm to compute the largest few thousands eigenvalues in the complex plane.   For the $G$ matrix of the diamond network (network III), about 1000 eigenvectors were computed. For $G^*$ matrix of diamond, about 500 eigenvectors were computed. 

Stochasticity of $G$ and $G^*$ implies that their spectra are necessarily inside the unit disk. For the World Wide Web the spectrum is spread inside the unit circle \cite{google}, with no gap between the largest eigenvalue and the bulk.
For networks I and II,  Fig.~\ref{spectrum} shows a huge gap between the first and the other eigenvalues. For the third network, there is still a gap between the first eigenvalue and next ones, but it is smaller.  While the distribution of the ranking vectors shown in Fig.~\ref{PR} reflects the distribution of links, the gap in the spectrum is related to the connectivity of the network and the presence of large isolated communities \cite{google}.  The presence of a large gap indicates a large connectivity, which is reasonable for the smaller networks. The presence of a smaller gap for network III indicates that there is more structure in the networks with larger plaquettes which disambiguate the different game paths and makes more visible the
communities of moves.  However, the gap being still present shows that even at the level of diamond-shaped plaquettes, the moves can belong to many different communities: this underlines one of the specificities of the game of go, which makes a given position part of many different strategic processes, and makes it so difficult to simulate by a computer.

The results in this Section show that the tools of complex networks such as ranking vectors associated to the largest eigenvalue already give new information which clearly go beyond the mere frequency count of the moves. This could be used to make more efficient the Monte Carlo algorithms of computer go. Nevertheless, other eigenvalues also carry valuable information, that we will study in the next Section.

\section{Eigenvectors and communities}

\begin{figure}[!h]
\begin{center}
\includegraphics[trim=0cm 0cm 0cm 0.0cm,clip,width=0.7\columnwidth]{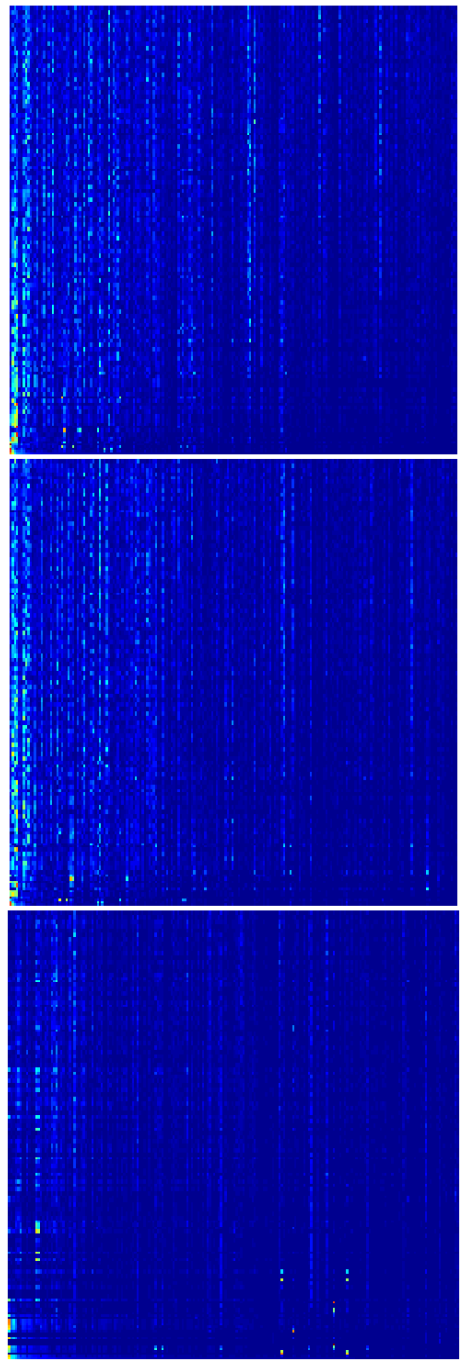}
\end{center}
\caption{(Color online) Eigenvector correlation map of the matrix $G$ for the three different networks : I (top), II (middle) and III (bottom). Top $200$ eigenvectors in order of decreasing eigenvalue modulus are plotted horizontally from bottom to top. Only the first 200 components are shown in the PageRank basis. The colors are proportional to the modulus of components (the normalization of an eigenstate $\psi$ is $\sum_i |\psi_i|^2=1$), from blue/dark grey (minimal) to red/light grey (maximal).\label{eigen}}
\end{figure}

In the preceding Section, we displayed the spectra of the networks constructed from the game of go. We have already discussed the ranking vectors associated to the largest eigenvalue. The other eigenvectors give a different information.
In Fig.~\ref{eigen} we display the intensities of the first 200 eigenvectors of the three different networks. It is clear that eigenvectors have specific features, not being spread out uniformly or localized around a single specific location. 
Correlations are also clearly visible between different eigenvectors, materialized by the vertical lines where several eigenvectors have similar intensities on the same node.  Correlations are less visible on the largest network, but it is also due to the much largest size of the vectors which decreases the individual projections on each node.  It is interesting to note that these correlations
are not necessarily
related to the PageRank values or the frequency of moves: vertical lines tend to be more visible on the left of the figure corresponding to high PageRank, but they are present all over
the interval: certain sequences of eigenvectors have correlated peaks at locations with relatively low PageRank.

\begin{figure}[!h]
\begin{center}
\includegraphics[trim=0cm 0cm 0cm 0.0cm,clip,width=0.9\columnwidth]{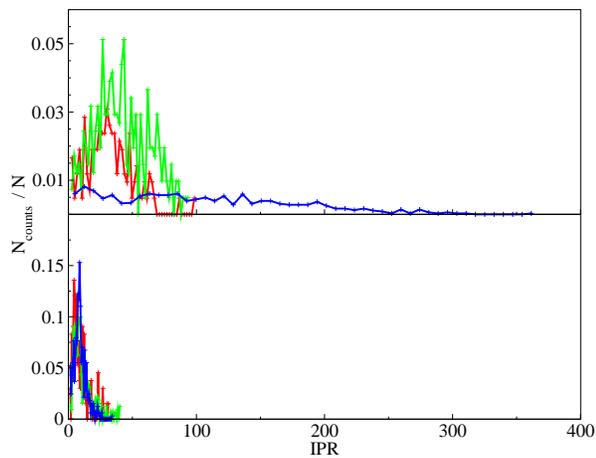}
\end{center}
\caption{(Color online) Histogram of IPR values (see text) for Network I (red/dark grey), Network II (green/light grey) and Network III (blue/black). Top panel shows the values
computed for eigenvectors of $G$ and bottom panel shows the same for $G^*$. Data correspond to the top $221$ eigenvalues (network I), top $410$ eigenvalues (network II) and top $999$  eigenvalues (network III).  \label{ipr}}
\end{figure}

In order to quantify these effects, we first look at the spreading of
eigenvectors: for a given vector, how many sites have significant projections? This can be measured for a vector $\psi$ through the Inverse Participation Ratio (IPR):
$\sum_i |\psi_i|^4/(\sum_i |\psi_i|^2)^2$. For a vector uniformly spread over
$P$ vertices it would be equal to $P$. A random vector thus has an IPR proportional
to the size of the system.
The data of Fig.~\ref{ipr} for the eigenvectors corresponding to the largest eigenvalues show that these vectors are not random or uniformly spread. On the contrary, their IPR is quite small, even for the largest network: in this case only a few dozen sites contribute to a given eigenvector, among almost 200000 possible nodes. Fig.~\ref{ipr} also shows that there is a relatively large dispersion of the IPR around the mean value. We provide the distributions for the Google matrices $G$ and $G^*$.  Qualitatively the  features are similar, but there is both a lower mean value and a  lower dispersion for $G^*$, indicating that the statistical symmetry found previously between incoming and outgoing links is indeed only approximate.

What is the meaning of these eigenvectors? If one interprets the Google matrix as describing a random walk among the nodes of the network as in the original paper \cite{brin}, eigenvectors of $G$ correspond to parts of the network where the random surfer gets stopped for some time before going elsewhere in the network. In other words, they are localized on sets of moves which are more linked together than with the rest of the network. This corresponds to so-called communities of nodes which share certain common properties. In social network, the importance of communities has been stressed several times and they are the subject of a large number of studies (see e.g. the review ~\cite{physrepcom}). The use of the eigenvectors of $G$ to extract the communities is one of the many available methods, which has been used already in the different context of the World Wide Web~\cite{comdima}. As already mentioned, eigenvectors with largest eigenvalues tend to be localized on groups of nodes where the probability is trapped for some time. This approach will thus detect communities of nodes from where it is difficult to escape, i.~e.~with few links leading to the outside. In parallel, the eigenvectors of $G^*$ tend to be localized on groups of nodes with few incoming links from the outside. Fig.~\ref{ipr} shows that this latter type of community, obtained from $G^*$, tends to be smaller on average for
the go game than the former type, obtained from $G$. These different communities should reflect different strategic groupings of moves during the course of the game.

\begin{figure}[!h]
\begin{center}
\includegraphics[trim=0cm 0cm 0cm 0.0cm,clip,width=0.9\columnwidth]{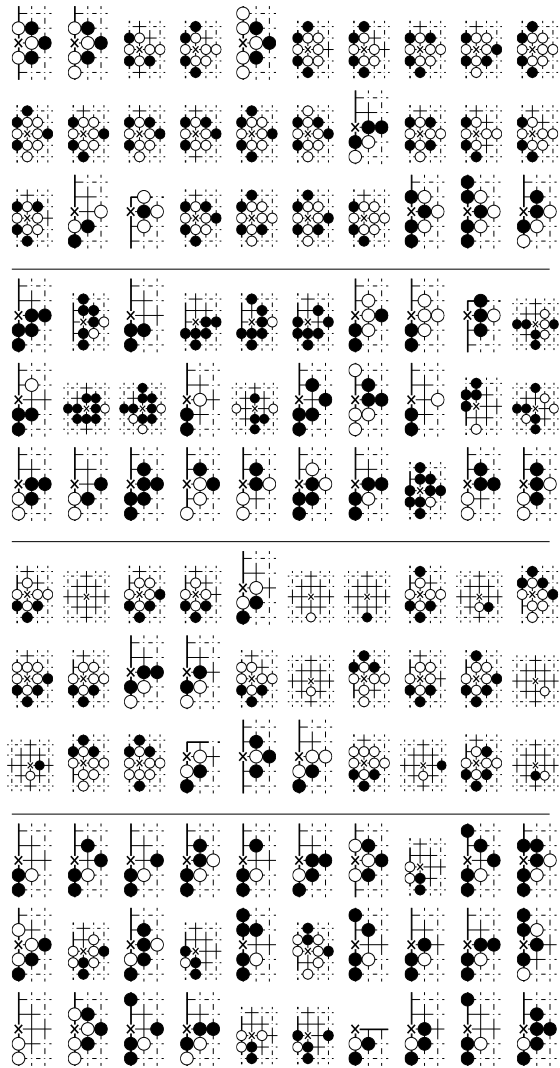}
\end{center}
\caption{Examples of the top $30$ nodes where eigenvectors of $G$ localize themselves for diamond network, from top to bottom $\lambda_{7}=-0.618$, $\lambda_{11}=0.185-0.5739i$, $\lambda_{13}=0.5651$, $\lambda_{21}=-0.4380$. \label{eigG}}
\end{figure}

\begin{figure}[!h]
\begin{center}
\includegraphics[trim=0cm 0cm 0cm 0.0cm,clip,width=0.9\columnwidth]{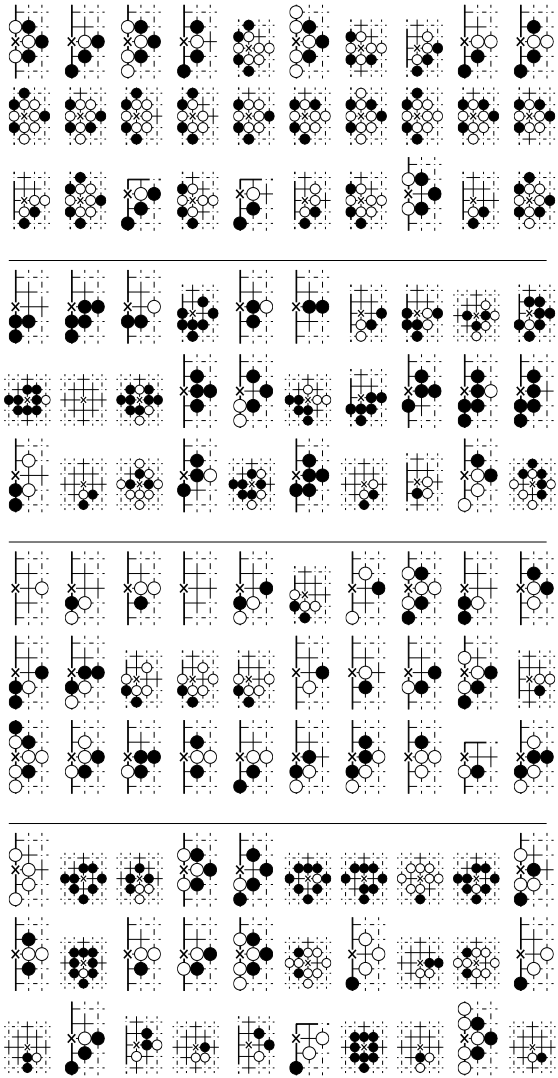}
\end{center}
\caption{Examples of the top $30$ nodes where eigenvectors of $G^*$ localize themselves for diamond network, from top to bottom $\lambda_{7}=-0.6023$, $\lambda_{11}=0.1743-0.5365i$, $\lambda_{18}=-0.4511$, $\lambda_{21}=-0.4021$. \label{eigG*}}
\end{figure}

The concept of community being intrinsically ambiguous, one can assign a subjective meaning to the definition of the community related to a chosen method. In our case, it is a difficult task to establish clear characteristics regarding what moves should be considered belonging to which community, however in the spirit of "moves that are more played together" or "similar moves" we can observe that a single eigenvector may contain a mixing of several communities. This could explain why in Fig.~\ref{eigen} one can see similar patterns appearing in different eigenvectors. These considerations are confirmed by Fig.~\ref{eigG} and Fig.~\ref{eigG*} where the first 30 moves of representative eigenvectors of $G$ and $G^*$ are displayed, ranked by decreasing component modulus.  While some common features appear, one gets the impression that groups of moves corresponding to different strategic processes are mixed and should be disentangled; for instance the last example of Fig.~\ref{eigG} seems to mix moves where black captures a white stone and moves where black connects a chain.

\begin{figure}[!h]
\begin{center}
\includegraphics[trim=0cm 0cm 0cm 0.0cm,clip,width=0.9\columnwidth]{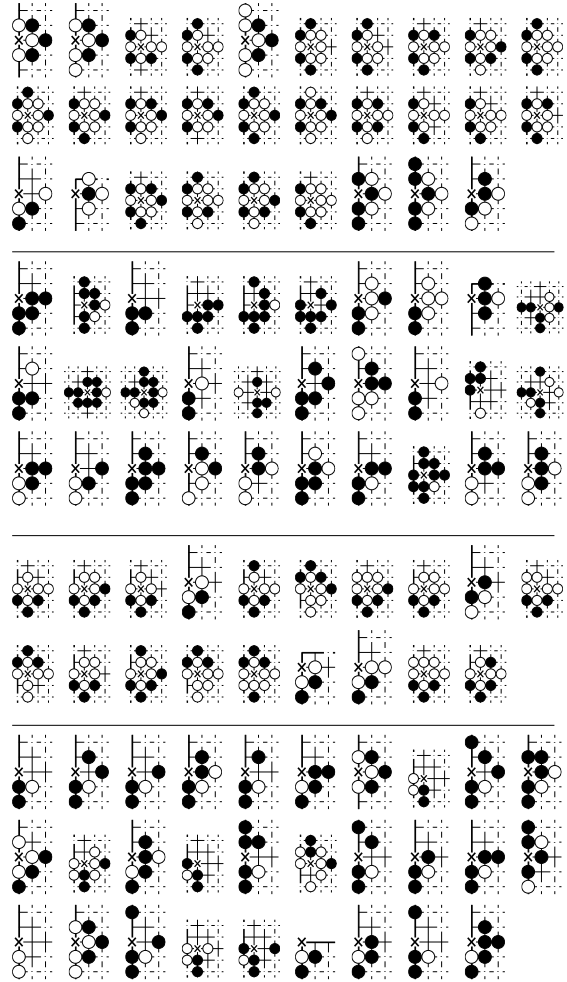}
\end{center}
\hskip 0.1cm
\caption{Same eigenvectors as in Fig.~\ref{eigG} treated by filtering out the top $30$ PageRank moves. \label{filterG}}
\end{figure}

\begin{figure}[!h]
\begin{center}
\includegraphics[trim=0cm 0cm 0cm 0.0cm,clip,width=0.9\columnwidth]{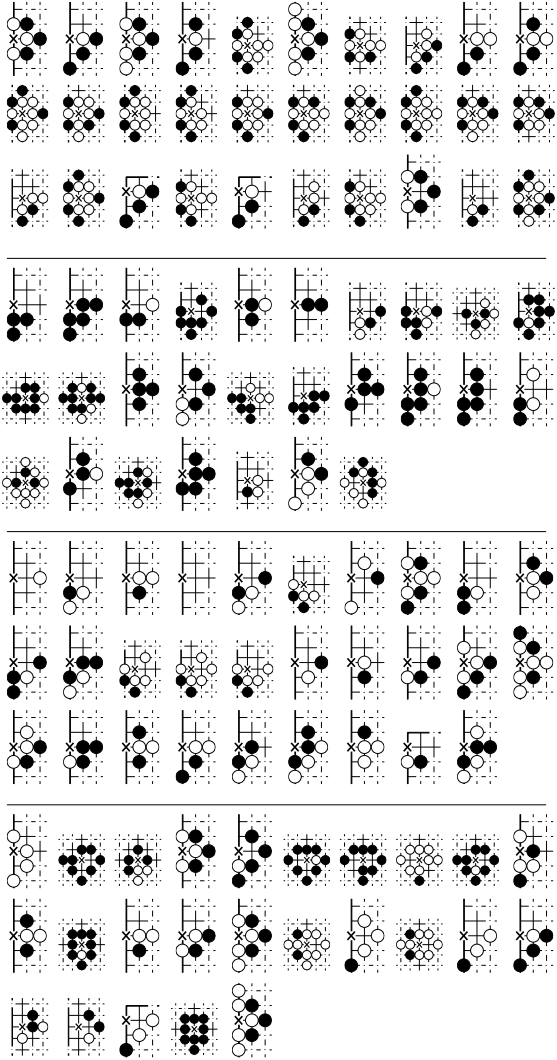}
\end{center}
\hskip 0.1cm
\caption{ Same eigenvectors as in Fig.~\ref{eigG*} treated by filtering out the top $30$ CheiRank moves. \label{filterG*}}
\end{figure}

In principle one could use correlations as the ones shown in Fig.~\ref{eigen} directly to identify communities, but we chose a different strategy. We propose here different basic methods that can be a first step into separating the communities within a given eigenvector. The simplest and most straightforward method consists in filtering out the effects of the most common and important moves by removing the top moves given by PageRank and CheiRank vectors. An example is shown in Fig.~\ref{filterG} and Fig.~\ref{filterG*} where the remaining moves in the given eigenvectors corresponds to a specific set of moves. Very common moves (such as empty or almost empty plaquettes) have been deleted, leaving more focused groups of moves. For example, the third eigenvector in Fig.~\ref{filterG} is much more focused on various moves containing situations of Ko or of imminent capture (Ko or ``eternity'' is a famous type of fights with alternate captures of opponent's stones).  

\begin{figure}[!h]
\begin{center}
\includegraphics[trim=0cm 0cm 0cm 0.0cm,clip,width=0.9\columnwidth]{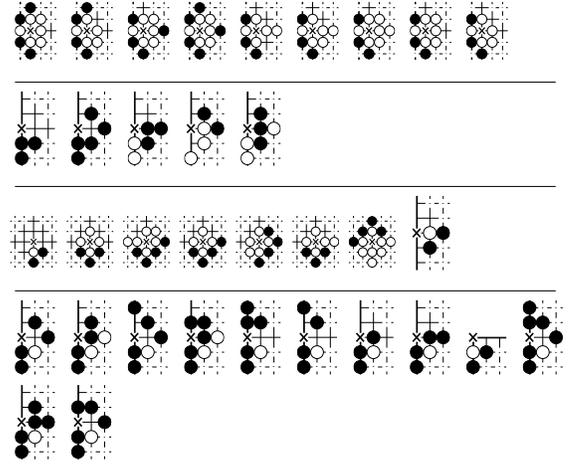}
\end{center}
\hskip 0.1cm
\caption{Example of set of moves extracted from data of Fig.~\ref{eigG} by considering common ancestry of moves with threshold level $\epsilon=0.3$ (see text) applied to $\lambda_{7}, \lambda_{11}$ and $\lambda_{21}$, and threshold level $\epsilon=0.5$ applied to $\lambda_{13}$. \label{ancestorG}.}
\end{figure}

\begin{figure}[!h]
\begin{center}
\includegraphics[trim=0cm 0cm 0cm 0.0cm,clip,width=0.9\columnwidth]{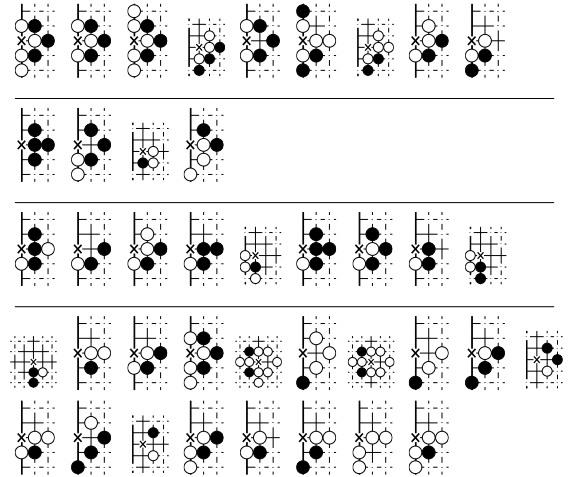}
\end{center}
\hskip 0.1cm
\caption{Example of set of moves extracted from data of Fig.~\ref{eigG*} by considering common ancestry of moves with threshold level $\epsilon=0.3$ (see text) applied to $\lambda_{7}, \lambda_{11}$, $\lambda_{18}$ and $\lambda_{21}$.\label{ancestorG*}}
\end{figure}

A more systematic method that we propose is to consider the ancestors of each move and determine if they share a significant number of preceding moves. As the Google matrix describes a Markovian transition model it would be natural to look for incoming flows of two moves to decide whether they belong to the same community. We implement it as follows: We choose two moves $m_1$ and $m_2$, with respectively $N_1$ and $N_2$ incoming links. We denote the origin of these incoming links pointing to $m_1$ and $m_2$ as sets of moves $S_1$ and $S_2$. If both moves share at least a certain fraction $\epsilon$ of common ancestors, that is if $\epsilon \; \mbox{min}(N_1,N_2) < \mbox{card} {(S_1 \cap S_2)}$, we assign both moves to the same community. This process is iterated until no more new moves are added to this community.
This extracting process is of course empirical, but helps us nevertheless to sort out some subgroups of moves that are different from those extracted with previous methods, provided that the parameter $\epsilon$ is carefully tuned. Indeed a too low value of $\epsilon$ does not help much in extracting a group as in most cases moves share naturally a certain amount of preceding moves but a too high value of $\epsilon$ will not capture anything for a sparse matrix.  In our Network III we thus used the range of values $0.3 < \epsilon < 0.7$. Unfortunately there is no typical behaviour of how the size of a community varies with respect to $\epsilon$: this size depends highly on the initial move and on the number of components of an eigenvector on which one is allowed to explore the ancestries.

We have applied this extracting process on eigenvectors. We thus identify communities in two steps, the first being to select eigenvectors corresponding to the largest eigenvalues of $G$ or $G*$, and the second step to follow this ancestry technique. As mentioned earlier an eigenvector corresponding to a large eigenvalue modulus is more likely to be localized on a small number of nodes, therefore one can truncate a given eigenvector to retain its top nodes and apply this method by choosing one of the top nodes as the starting move and constructing the community by successively exploring this subset. Starting from different nodes will allow to identify the different communities.
  Fig.~\ref{ancestorG} and Fig.~\ref{ancestorG*} show that the method is able to extract moves which have common features, much more so that just looking at largest components of the vectors or removing the ranking vectors (as in Fig.~\ref{eigG}, Fig.~\ref{eigG*}, Fig.~\ref{filterG} and Fig.~\ref{filterG*}). Small subsets of moves are disambiguated from the larger groups
of the preceding figures, showing sequences which seem to go together with situations of Ko with different black dispositions (first and third eigenvector of Fig.~\ref{ancestorG}), black connecting on the side of the board (fourth eigenvector of Fig.~\ref{ancestorG}), and so on.  Similarly, the first line of Fig.~\ref{ancestorG*} can be associated to attempts by black to take over an opponent's chain on the rim of the board. These examples show that the method is effective to regroup moves according to reasonably defined affinities.  

\begin{figure}[!h]
\begin{center}
\includegraphics[trim=0cm 0cm 0cm 0.0cm,clip,width=0.9\columnwidth]{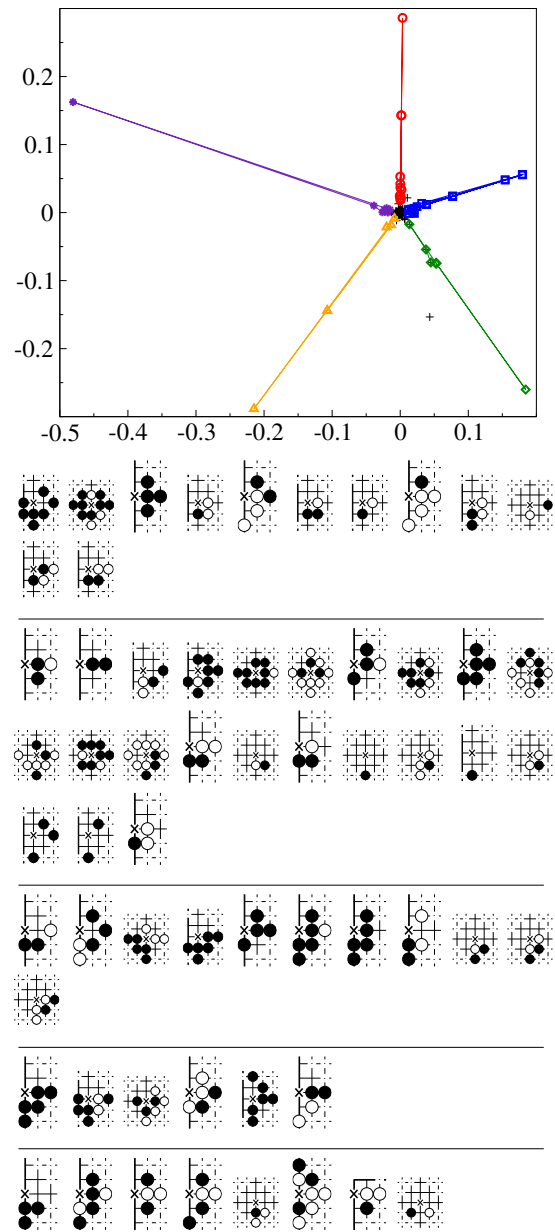}
\end{center}
\hskip 0.1cm
\caption{(Color online) Example of community extraction through phase analysis (see text) applied on the eigenvector $\psi$ of $G*$ corresponding to $\lambda_{13}$. Top: eigenvector components in the complex plane; groups of plaquettes, from top to bottom, correspond to respective symbols red circles, blue squares, green diamonds, oranges triangles and purple stars. \label{phaseanalysis}}
\end{figure}
          
We mention an alternative method which gives good results in some instances. It consists in analyzing the angles of an eigenvector components when plotted in a complex plane. This method is not systematic as there exist several real valued eigenvectors but for the complex ones one can observe interesting patterns. Either the plots show a meaningless cloud of points or they can reveal a tendency of a subset of components to be aligned. As shown in an example in Fig.~\ref{phaseanalysis} there can be one or several directions within the same eigenvector, indicating that maybe the phases of the components can characterize moves sharing common properties. Qualitatively speaking the spatial configuration of these subgroups of moves look similar but there are also similarities between moves having different angles, and a formal understanding of the meaning of phases is still lacking. We note that for undirected networks the sign of components of eigenvectors of the adjacency matrix has been used to detect communities \cite{krz}.

It is worth insisting again on the fact that in general the next to leading eigenvectors in the Google matrix represent a different information from the list of most common moves. In fact, these eigenvectors can even sometimes be highly sensitive to rare links, indeed during our analysis one impossible move was highlighted in one of the top eigenvectors. This move had only two links among the several millions, leading us to find a fake gamefile in the dataset.  This shows that the network approach can detect specificities that a mere statistical analysis of the datasets will miss. 

It is in principle not excluded that one should look into combinations of eigenvectors but even though we considered single vectors, the results show that it is possible to extract community of moves which share some common properties with these methods. The combination of methods outlined in this section, namely isolating top moves in eigenvectors associated to large eigenvalues, and disambiguating them through search for common ancestries, seems to yield meaningful groups of moves. We stress again that they do not merely correspond to most played moves or sequences of moves, nor to the best ranked in the PageRank or CheiRank, but give a different information related to the network structure around these moves. It is possible to play with the parameters of the method (threshold $\epsilon$, number of eigenvectors, starting point of the common ancestry) in order to find different sets of communities, which should be analyzed in relation with the strategy of the game, and then could help organize the Monte Carlo go search by running it into specific communities.

\section{Generalized networks}

One can refine the analysis further by disaggregating the datasets in several ways, constructing different networks from the same database. The number of nodes is still the same, but links are now selected according to some specific criterion and may give rise to different properties.  In this Section we will illustrate this by a few examples.

\begin{figure}[!h]
\begin{center}
\includegraphics[trim=0cm 0cm 0cm 0.0cm,clip,width=0.9\columnwidth]{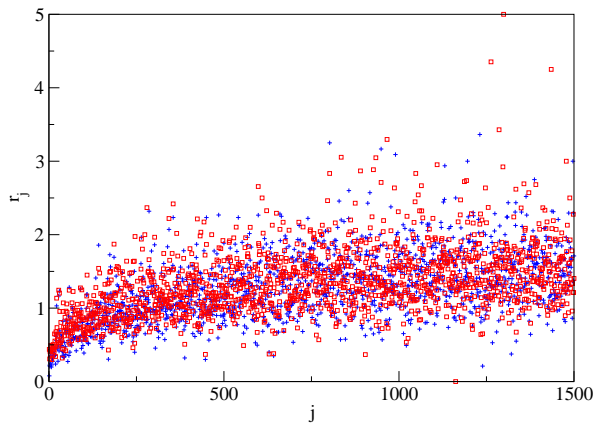}
\end{center}
\caption{(Color online) Fluctuation difference $r_j=\sum_{i \leftarrow j} |k_i - k'_i|/\sum_i k_i$ of outgoing links versus move indices for top 1500 moves of diamond patterns in PageRank order (network III)(see text). An example of difference is shown between two networks built from games between 6d players (blue crosses) and two networks built respectively from games between 1d players and games between 9d players (red squares). The number of games in each case is 2731, corresponding to the number of 1d/1d games in the database \cite{database}. \label{rt}}
\end{figure}

\begin{figure}[!h]
\begin{center}
\includegraphics[trim=0cm 0cm 0cm 0.0cm,clip,width=0.9\columnwidth]{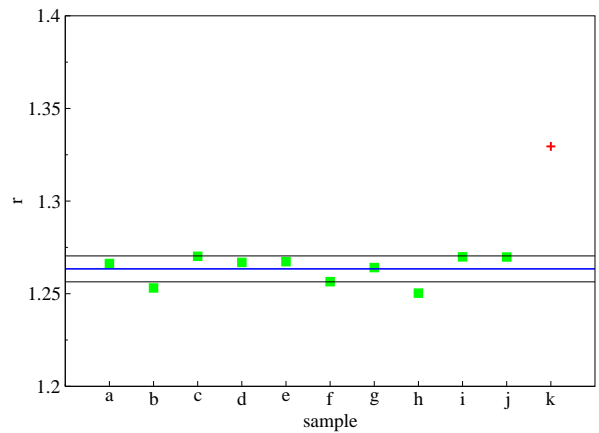}
\end{center}
\caption{(Color online) Difference $r$ (see text) between the networks built from games of 1d players and of 9d players (red cross) together with several examples of $r$ for pairs of networks constructed from different samples of games of 6d players (green squares). The three horizontal lines mark the mean and the variance of the 6d values  The number of games in each sample is 2731, corresponding to the number of 1d/1d games in the database.  \label{histo}}
\end{figure}

An important aspect of the games, especially in view of applications to computer go, is to select moves which are more susceptible of winning the game. It is possible to separate the players between winners and losers, but the presence of handicaps makes this process ambiguous. Indeed, it is possible to place up to nine stones before the beginning of the game at strategic locations, giving an advantage to a weaker player which may allow him to play against a better opponent with a fair chance of winning. Another possibility we thus investigated was to separate the players by their levels according to their dan ranking. Indeed, players are ranked from first dan (1d, lowest level) to ninth dan (9d, highest level).  In the database \cite{database} the number of dans of the players is known, and it is therefore possible to separate games played at different levels. To explore these differences, we constructed the diamond network from games played by 1d versus 1d, the one from 9d versus 9d, and the one from 6d versus 6d.  Fig.~\ref{rt} shows the quantity $r_j=\sum_{i \leftarrow j} |k_i - k'_i| /\sum_i k_i$ defined for a pair of networks, where $k_i$ (resp. $k'_i$) is the number of links from a fixed node $j$ to node $i$ for one network (resp. for the second network). For each node, $r_j$ thus quantifies the difference in outgoing links between two networks.  Fig~\ref{rt} shows the distribution of this quantity highlighting the difference between the network 1d/1d and the network 9d/9d. One sees that they are indeed different, with a mean $\langle r_j \rangle \approx 1.33$. Nevertheless, in the same figure we add for comparison the difference between two networks of 6d/6d, showing that one can also find differences between networks built from players of the same level.  In view of this, to see if the difference between 1d/1d and 9d/9d is statistically significant, Fig~\ref{histo} shows the average $r= \langle r_j \rangle$ for different choices of samples of 6d versus 6d games and the value for the networks constructed from the games of 1d players and 9d players, with the average taken on top $1500$ moves of the PageRank. It shows that the difference between 1d players an 9d players has some statistical significance. The quantity $r$ is a simple way of quantifying the structural differences in the networks at the level of outgoing flows which is in our case an indication that 9d players might have an overall structurally different style of play than 1d players, even though the difference is relatively small.

\begin{figure}[!h]
\begin{center}
\includegraphics[trim=0cm 0cm 0cm 0.0cm,clip,width=0.9\columnwidth]{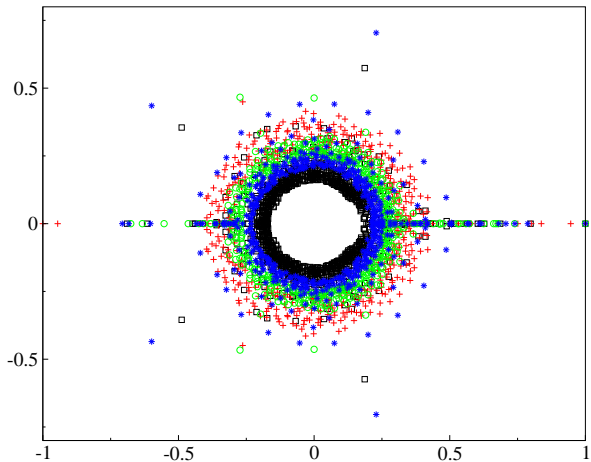}
\end{center}
\caption{(Color online) Spectrum of $G$ for diamond networks of different game phases : first $50$ moves (red crosses), middle $50$ moves (green circles) and last $50$ moves (blue stars). The black squares correspond to the spectrum of the network when the whole game is taken into account, shown for reference.\label{phases}}
\end{figure}

An interesting possibility which might also be useful for applications is to create separate networks for different phases of the game. For instance, one can take into account when using the database of real games only the first $50$ moves, the middle $50$, or the final $50$. Again, this does not modify the nodes of the networks, but changes the links, creating three different networks corresponding to respectively beginning, middle, and ending phases of the game. The number of links is now 6155936 for the beginning phase, 6460771 for the middle phase, and 5947467 for the ending phase (instead of 26116006 for the whole game) (the numbers without degeneracies for diamond plaquettes are respectively 613953, 2070305 and 3182771). The spectra of the three networks for the diamond plaquettes are shown in Fig.~\ref{phases} (again, only the largest eigenvalues are calculated). It is clear that the spectra are quite different, indicating that the structure of the network is not equivalent for the different phases of the game. It is visible that the eigenvalue cloud is larger for the ending phase indicating that near the final stage of the game the random surfer gets trapped more easily in specific patterns, which should correspond to typical endgames. Similarly, the gap is smaller for the beginning phase, indicating that one strongly knit community exists with an eigenvalue close to the PageRank value.

\begin{figure}[!h]
\begin{center}
\includegraphics[trim=0cm 0cm 0cm 0.0cm,clip,width=0.9\columnwidth]{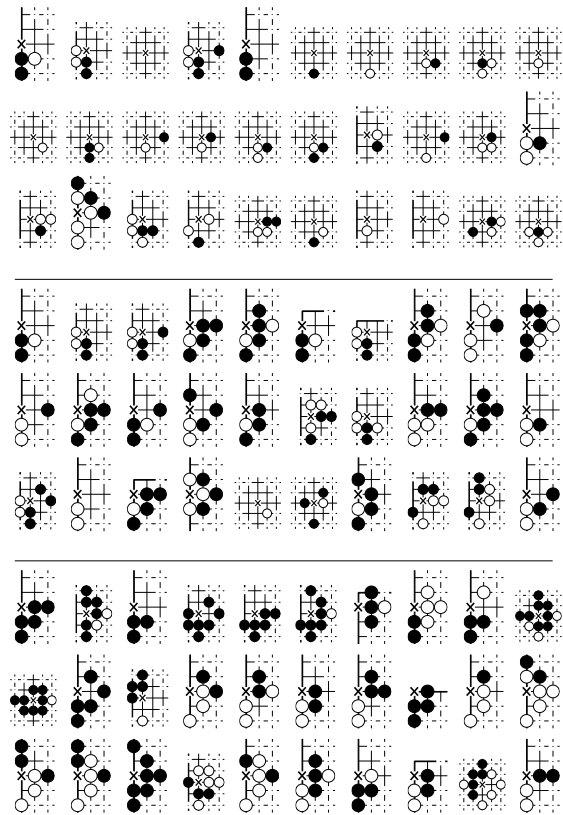}
\end{center}
\caption{Examples of set of top $30$ moves where eigenvectors localize themselves, those examples are computed for diamond network in different game phases : starting phase and $\lambda_{4}$ (top), middle phase and $\lambda_{4}$ (middle) and ending phase $\lambda_{4}$ (bottom). \label{gamephase}}
\end{figure}

The eigenvectors shown in Fig.~\ref{gamephase} highlight different sets of moves as might be expected since strategy should differ in those phases. Obviously, eigenvectors for opening moves are much more biased towards relatively empty plaquettes, indicating the start of local fights. In the middle and end of the games, communities are biased towards moves corresponding to more and more filled plaquettes,  indicating ongoing fights or fight endings. We stress the fact that those sets of moves are not just the most played moves in the respective phases. Running the community detection process of Section IV on such eigenvectors should select communities specific to these different phases of the game.

\section{conclusion}

We have shown that it is possible to construct networks which describe the game of go, in a spirit similar to the ones already used for languages. We have extended the results of \cite{gonet}, comparing three networks of different sizes according to the size of the plaquettes which serve as nodes of the network. The three networks share structural similarities, such as a statistical correlation (but not an exact symmetry) between incoming and outgoing links. However, the largest network, besides necessitating more refined numerical tools in order to obtain the largest eigenvalues and associated eigenvectors, is also much less connected and disambiguates much better the different moves. We have also shown that specific subnetworks can be constructed, selecting links in the databases according to levels of the players or phases of the game.

Our results show that the networks constructed in this way have specific properties which reflect the peculiarities of the game. In particular, the PageRank and CheiRank vectors give new orderings of the moves, which do not merely correspond to most played moves or sequences of moves, but give a different information. As explained in Section III, moves highlighted by the ranking vectors can correspond to moves which are connected to chains of important moves, eventhough they are not that frequent (it was this difference which made Google the famous company it is today).  We have also shown that it is possible with these methods to extract communities of moves which share some common properties. A possible use of these results would be to help organize the Monte Carlo go search by running it into specific communities.
Indeed, despite its limitations \cite{HuaMul13}, Monte-Carlo go remains the most promising approach to computer go. The main goal of these algorithms is an efficient value function estimation \cite{GelKoc12}. We have proposed in this paper various community detection processes, and the knowledge of these communities could be used for instance to initialize the value of moves according to the local pattern, at a value given by the value of its ancestors. It could also be used to propagate the value of a move to similar moves. It would be interesting to compare the values assigned to nodes of our networks by the different computer programs available, in order to see whether adjacency matrix properties could be used to converge more quickly to the correct value function.  We think an especially interesting path in this direction corresponds to the approach outlined in Section V: by constructing specific networks according to game phases or levels of players, one can specify communities useful in specific contexts of the game or corresponding to winning strategies.  It is also possible to use ``personalization''techniques (implemented by modifying the vector $e$ in the definition of $G$ in Section III \cite{googlebook}) which are currently explored in a World Wide Web context and allow to compute a ranking vector biased towards a certain group of nodes, e.~g.~one of the communities discussed in Section IV. All these techniques deserve further study in this context.

It will be fascinating to see if other games such as chess could be modelized this way, and how different the results will be.  Besides its applicability to the simulations of go on computers, we also believe that such studies enable to get insight on the way the human brain participates in such game activities, and more generally on the human decision-making processes \cite{Brown12}. In this direction, an interesting extension of this work could be to compare the networks built from games played by human beings and computers, and determine how different they are.

\begin{acknowledgments} 

We thank Dima Shepelyansky, Klaus Frahm, Pierre Aubourg, Yoann S\'eon and Fran\c{c}ois Damon for discussions and insights. We thank CalMiP for access to its supercomputers. V. K. thanks the CNRS and the R\'egion Midi-Pyr\'en\'ees for funding. This research is supported in part by the EC FET Open
project ``New tools and algorithms for directed network
analysis'' (NADINE No 288956).

\end{acknowledgments}
 
%\clearpage

\end{document}